\documentclass[a4paper,11pt]{article}
\usepackage{pos}

\newcommand{\hess}{HESS\,J1023-575}
\newcommand{\psr}{PSR\,J1023--5746}
\newcommand{\ext}{FGES\,J1023.3--5747}
\title{Probing the hadronic nature of the gamma-ray emission associated with Westerlund 2}

\author*[a]{Emma de O\~na Wilhelmi}
\author[b,c]{Enrique Mestre}
\author[b,c,d]{Diego F. Torres}
\author[e]{Tim Lukas Holch}
\author[e]{Ullrich Schwanke}
\author[f,g,h]{Felix~Aharonian}
\author[i,j]{Pablo Saz Parkinson}
\author[k,l,m]{Ruizhi Yang}
\author[n]{Roberta Zanin}

\affiliation[a]{Deutsches Elektronen Synchrotron DESY, 15738 Zeuthen, Germany}
\affiliation[b]{Institute of Space Sciences (ICE/CSIC), Campus UAB, Carrer de Can Magrans s/n, 08193 Barcelona}
\affiliation[c]{Institut d'Estudis Espacials de Catalunya (IEEC), 08034 Barcelona, Spain}
\affiliation[d]{Instituci\'o Catalana de Recerca i Estudis Avan\c cats (ICREA), E-08010 Barcelona, Spain}
\affiliation[e]{Humboldt University of Berlin, Newtonstr. 15, 12489 Berlin, Germany}
\affiliation[f]{Dublin Institute for Advanced Studies, 31 Fitzwilliam Place, Dublin 2, Ireland}
\affiliation[g]{Max-Planck-Institut f\"ur Kernphysik, P.O. Box 103980, D 69029 Heidelberg, Germany}
\affiliation[h]{Gran Sasso Science Institute, 7 viale Francesco Crispi, 67100 L'Aquila,  Italy}
\affiliation[i]{Department of Physics and Laboratory for Space Research, The University of Hong Kong, Pokfulam Road, Hong Kong}
\affiliation[j]{Santa Cruz Institute for Particle Physics, University of California, Santa Cruz, CA, 95064, USA}
\affiliation[k]{Department of Astronomy, School of Physical Sciences, University of Science and Technology of China, Hefei, Anhui 230026, China}
\affiliation[l]{CAS Key Labrotory for Research in Galaxies and Cosmology, University of Science and Technology of China, Hefei, Anhui 230026, China}
\affiliation[m]{School of Astronomy and Space Science, University of Science and Technology of China, Hefei, Anhui 230026, China}
\affiliation[n]{CTA Observatory GmbH, Via Piero Gobetti 93, I-40129 Bologna, Italy}

\emailAdd{emma.de.ona.wilhelmi@desy.de}
\emailAdd{mestre@ice.csic.es}

\abstract{Star-forming regions have been proposed as potential Galactic cosmic-ray accelerators for decades. Cosmic ray acceleration can be probed through observations of gamma-rays produced in inelastic proton-proton collisions, at GeV and TeV energies. We analyze more than 11 years of {\it Fermi}-LAT data from the direction of Westerlund 2, one of the most massive and best-studied star-forming regions in our Galaxy. The spectral and morphological characteristics of the LAT source agree with the ones in the TeV regime (HESS J1023-575), allowing the description of the gamma-ray source from a few hundreds of MeV to a few tens of TeVs. We will present the results and discuss the implications of the identification with the stellar cluster and the radiation mechanism involved.}

\FullConference{%
  *** 37th International Cosmic Ray Conference (ICRC2021), ***\\
  *** 12-23 July 2021 ***\\
  *** Berlin, Germany - Online ***
}

\begin{document}
\maketitle

\section{Introduction}

The potential of massive star clusters to accelerate Galactic cosmic rays (GCRs) to very-high energies (VHE, E > 100 GeV) has been recognized since the 1980s \citep{1983SSRv...36..173C,2020SSRv..216...42B,2005ApJ...634..351B}. Several hypotheses have been proposed for acceleration sites to very high energies in star-forming regions (SFRs), either in the vicinity of OB and WR stars, or at the interaction of their fast winds with supernova (SN) shocks, or in so-called super-bubbles. The presence of cosmic rays (CRs) can be inferred through gamma-ray observations (above a few hundreds of MeV), by looking at the by-product of inelastic proton-proton collisions. The spectral energy distribution (SED) from hadronic-originated gamma-ray sources is characterized by a sharp rise in the $\sim$70--200 MeV range (resulting from the neutral pion threshold production energy), followed by a hard emission up to the maximum energy. Competing gamma-ray processes related to leptonic emission should have different imprints on the spectral shape, showing a (in many occasions broad) peak in the hundreds of GeV range. Therefore, sampling the spectrum from a few hundreds of MeV to tens of TeV should result in a strong indication of the origin of the observed radiation. 

We investigated the spectral and morphological characteristics of the {\it Fermi}-LAT source FGES\,J1023.3--5747, which is spatially in agreement with one of the most massive and well-studied SFRs in our Galaxy: Westerlund\,2 \cite{2021MNRAS.tmp.1423M}. The cluster, which contains an extraordinary ensemble of hot and massive OB stars \citep{2004A&A...420L...9R}, has been previously associated with an extended TeV source ($\sim$0.2$^{\rm o}$) known as HESS\,J1023-575 \citep{2007A&A...467.1075A,2011A&A...525A..46H}. However, a bright GeV pulsar, dubbed PSR\,J1023--5746, was also found in the vicinity of the TeV source, 8\,arcmin away from the cluster. Given the efficiency of PWNe to produce gamma-ray in the TeV regime \citep{2018A&A...612A...2H}, a possible identification of the extended GeV and TeV emission with the PWN powered by the pulsar has been also suggested \citep{2010ApJ...725..571S,2011ApJ...726...35A}.

 The detailed analysis of 11 years of \emph{Fermi}-LAT, using updated ephemeris to gate off the gamma-ray emission from the pulsar, allows the investigation of the spectral shape in the 200 MeV to 100 GeV energy range, as well as the morphology of the GeV source. The combined results provide new hints on the origin of the GeV and TeV emission and its connection with Westerlund 2. 

\section{Results}
To investigate the characteristics of the gamma-ray emission, we used \emph{Fermi}-LAT (P8R3, \citep{2013arXiv1303.3514A,2018arXiv181011394B}) data spanning from 2008 August 4 to 2019 April 24 (or 239557417 - 577782027 seconds in \emph{Fermi} Mission Elapsed Time), in the energy range from 200\,MeV to 500\,GeV. We retrieved the data from a region of interest (ROI) defined by a radius of 20$^{\rm o}$ around the position of \psr\ (RA= 155.76$^{\rm o}$, DEC = -57.77$^{\rm o}$, \cite{2015ApJ...814..128K}). The events were selected with a minimum energy of 200\,MeV, to avoid events poorly reconstructed due to the large angular resolution and the large crowding of sources in the region. The analysis of the LAT data described above was performed by means of the \textsc{fermipy} \textsc{python} package (version 0.18.0), based on the \textsc{Fermi Science Tools} \citep{2017ICRC...35..824W}. More details on the analysis are described in \cite{2021MNRAS.tmp.1423M}.

\begin{figure}
  \centering
  \includegraphics[width=0.5\textwidth]{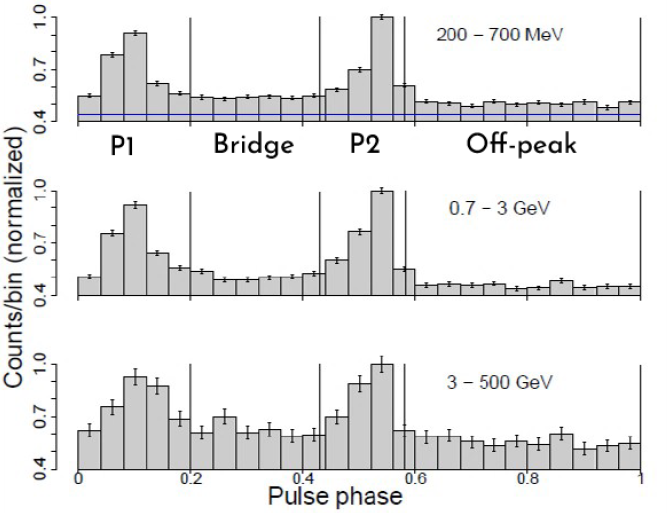}
  \caption{Phase curve of PSR J1023.1-5745 obtained at different energies in a region of 0.6$^{\rm o}$ around the pulsar position (normalized by the height of P2) with the different components of the phase curve noted.
  The horizontal blue line corresponds to the expected contribution in the phase curve of background sources (i.e., all the sources in the field of view except for \ext\ and \psr). The light curve was obtained in three intervals of energy (from top to bottom): from 200\,MeV to 700\,MeV, from 700\,MeV to 3\,GeV, and above 3\,GeV.}
  \label{fig:phasecurve1023}
\end{figure}
\subsection{Pulsar gating-off}

We performed a timing analysis to gate the pulsed emission, using the pulsar analysis package \textsc{TEMPO2} \citep{2006MNRAS.369..655H}. To obtain the phase curve of \psr, we assigned the corresponding phases to the gamma-ray events localized in a region of 0.6$^{\rm o}$ of radius around the pulsar position (RA = 155.76$^{\rm o}$, DEC = -57.77$^{\rm o}$), using the updated ephemeris of \psr\ at epoch 55635 MJD. The phaseogram is shown in Fig. \ref{fig:phasecurve1023}. Applying the Bayesian Blocks method \citep{2013arXiv1304.2818S}, we derived different components of the light curve, defined as ON, OFF, and {\it Bridge} emission. We computed the contribution to the phase curve of each of the sources in the model with the \emph{Fermi} tool {\it gtsrcprob}, which assigns to every event the probability of belonging to each source of the model. The second peak dominates the light-curve in the three intervals of energy shown in Fig. \ref{fig:phasecurve1023}. The number of events in the {\it Bridge} component is $\sim 10$ percent larger than the one expected from the off-peak statistics. It should be noted that both the {\it Bridge} and off-peak regions are dominated by the Galactic diffuse emission (which results in a similar number of count for both components). This introduces important systematic errors, especially in the lower energies, when investigating the gamma-ray emission associated to \ext\ (see below). 

To investigate the emission associated with the extended source \ext, we disregard events outside the off-peak interval (from 0.58 to 1 in phase), minimizing so the contamination from \psr.

\subsection{The extended source \ext}

The spectral-morphological analysis of \ext\ in the off-peak region results on a significance of ($\sqrt{TS} \approx 14$). The source spatial component is best described by a symmetric 2D Gaussian model centred at RA= 155.93$^{\rm o}\pm 0.03^{\rm o}$ and DEC = -57.79$^{\rm o} \pm 0.03^{\rm o}$, with an extension, in a 68 percent containment radius, of $r_{68} = 0.24^{\rm o}\pm 0.03^{\rm o}$ (i.e., an intrinsic size of $\sigma = 0.16 \pm 0.02$). These values are in very good agreement with the one measured by H.E.S.S. on the TeV source \hess, with an extension of $\sigma_{\rm TeV} = 0.18 \pm 0.02$ and located $3.3 \pm 1.4$ arcmin away \citep{2007A&A...467.1075A,2011A&A...525A..46H}. The spectral energy distribution is well described by a power-law function, normalised to 17 GeV flux of $N_{0}$ = $(1.02 \pm 0.14_{stat} \pm 0.16_{sys}) \times 10^{-14}$ $\rm{cm}^{-2} \rm{s}^{-1} \rm{MeV}^{-1} \rm{sr}^{-1}$, and spectral index $\Gamma$ = $2.05 \pm 0.06_{stat} \pm 0.33_{sys}$.

\subsection{Systematic studies}

\begin{figure}
  \centering
  \includegraphics[width=0.38\textwidth]{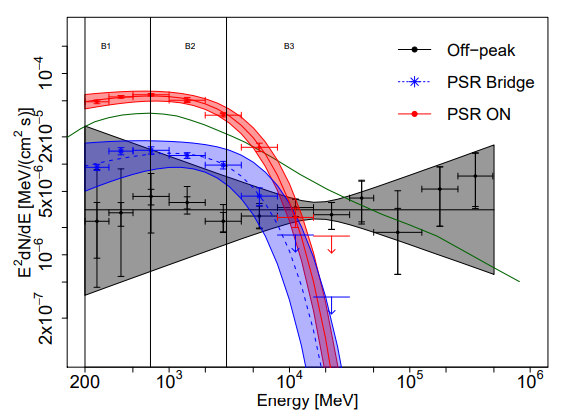}
  \includegraphics[width=0.57\textwidth]{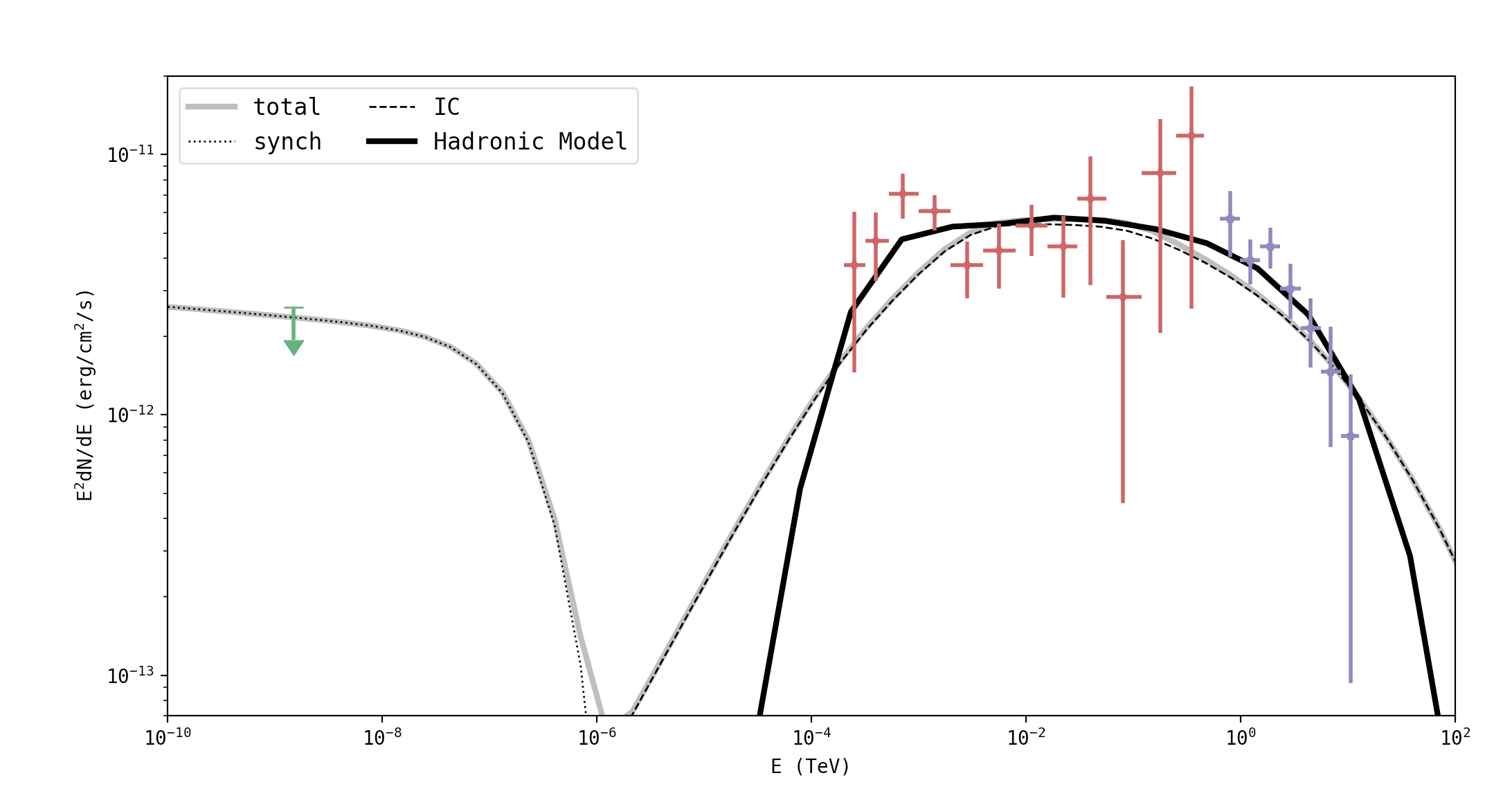}
  \caption{{\bf Left:} Spectral energy distribution of \psr\ derived from the on-peak (in red) the {\it Bridge} (in blue), and off-peak region (in black). {\bf Right:} Best-fitting models for \ext\ spectrum in the pion decay (solid line) and PWN (dashed line) hypotheses (see \citep{2021MNRAS.tmp.1423M} for details on the modeling). The red points correspond to the \emph{Fermi}-LAT data analysis (with only statistical errors), and the purple ones to H.E.S.S.  \citep{2011A&A...525A..46H}. The upper limit in the X-ray domain is obtained from \citep{2009PASJ...61.1229F}.}
  \label{fig:systematics}
\end{figure}

Given the extension and flux level of \ext\ a large contamination of the overwhelming diffuse Galactic emission is expected. To assess its contribution, we compare the expected gamma-ray emission in a region of the size of \ext, with the spectra derived from the pulsar in the ON-peak and {\it Bridge} region, and the extended source underlying (OFF-region). The results are shown in Fig. \ref{fig:systematics}, with the ON-peak and {\it Bridge} in red and blue respectively, the LAT emission associated to \hess\ in black (OFF-peak), and the Galactic diffuse emission in green. The energy intervals labeled as B1, B2, and B3 correspond to the energy ranges in Fig.\ref{fig:phasecurve1023}. The stability of the results was investigated by looking at the ratio between the integral flux of each component and the expected contribution of the diffuse galactic emission, in each of the light-curve regions. These ratios are shown in Tab.1.

\begin{table}[htbp]
\caption{Flux ratio of the different components with respect to the Galactic diffuse emission}
\label{tab:ratio}
\resizebox{\columnwidth}{!}{\begin{tabular}{|c|c|c|c|c|c|c|}

 \hline
Energy range  & \multicolumn{2}{|c|}{On-Peak} & \multicolumn{2}{|c|}{{\it Bridge}} & \multicolumn{2}{|c|}{Off-Peak}\\

 &  PSR J1023-5746 & \ext & PSR J1023-5746 & \ext & PSR J1023-5746 & \ext \\
 \hline
 B1 & 1.94$^{+0.27}_{-0.25}$ & 0.27$^{+0.2}_{-0.11}$ & 0.33$^{+0.14}_{-0.1}$ & 0.18$^{+0.15}_{-0.08}$ & --  & 0.11$\pm0.03$ \\
 B2 & 1.98$^{+0.21}_{-0.19}$ & 0.22$^{+0.11}_{-0.07}$ & 0.38$\pm0.09$ & 0.15$^{+0.09}_{-0.05}$ & --  & 0.1$\pm0.02$ \\
  B3 & 1.28$^{+0.28}_{-0.23}$ & 0.36$^{+0.10}_{-0.09}$ & 0.3$^{+0.14}_{-0.12}$ & 0.28$^{+0.08}_{-0.07}$ & --  & 0.21$^{+0.04}_{-0.03}$ \\
 \hline

\end{tabular}}
\end{table}

The integrated flux of \ext\ is compatible (at 95\% CL) between components (within each energy range). The statistical error increases, as expected, in the On-peak and {\it Bridge} region, due the contamination of the pulsar. The integral flux from the Galactic
diffuse emission, given an energy interval, remains constant for the different components, pointing to a good estimation of the spectrum, even if the emission is dominated by the diffuse emission in the Off and {\it Bridge} regions.

To further investigate the systematic uncertainties on the spectral, we artificially modify the normalization of the diffuse Galactic model by $\pm 6$ percent with respect to the best-fitted value \citep{2012ApJ...744...80A,2018ApJ...858...84L}. Additionally, systematic uncertainties due to the LAT effective area are computed with the bracketing $\rm{A}_{eff}$ method\footnote{\url{https://fermi.gsfc.nasa.gov/ssc/data/analysis/scitools/Aeff_Systematics.html}}. The errors were added quadratically to the statistical ones and displayed in the spectrum in Fig. \ref{fig:systematics}.

\section{Westerlund\,2 as a power house of \ext}

The matching morphology and smooth spectral connection between the LAT and H.E.S.S. sources allow us to firmly identify \ext\ as the GeV counterpart of \hess. The lack of energy-dependent morphology, expected in a leptonic scenario with continuous injection of particles (as in a PWN), and the extension of the hard spectrum into a few hundreds of MeV, can be explained naturally in a hadronic scenario. In such scenario, energetic CRs interact with molecular material in the region, originating the observed gamma-ray spectrum. To constrain the proton population that powers the gamma-ray source we model the 200 MeV to 20 TeV emission using the {\sc naima} package (version 0.8.4, \citep{2017ascl.soft08022Z}). To calculate the SED, we used a distance of 5 kpc. The molecular content in the region has been deeply investigated by several authors \citep{2007ApJ...665L.163D,2014ApJ...781...70F} using millimeter-wave CO spectroscopy. Several massive molecular clouds were found within the surrounding of Westerlund 2. The total mass is estimated to be between $(1.7-7.5)\times10^{5}M_{\odot}$. We used a particle distribution described by a particle index $s$ and an amplitude $N_{p}$, up to an energy cut-off $\rm{E}_{cutoff}$. The corresponding gamma-ray emission due to pion decay radiation is calculated and compared to the experimental data. 
The best-fitting model for the joint \emph{Fermi}-LAT and H.E.S.S. data corresponds to an exponential cut-off power-law proton spectrum with cut-off energy $\rm{E}_{cutoff} = 93 \pm 8$ TeV, and a particle index of $s = 2.09 \pm 0.01$, referenced to 1\,TeV, 
see Fig. \ref{fig:systematics}). For a distance of 5 kpc, the total energy in protons estimated above 1.22\,GeV (the threshold kinetic energy for pion production in pp interactions) is $W_{\rm p} = (1.3 - 5.9)\times10^{48}$ erg, for densities of n = (7.5 - 1.7)$\times10^5$ M$_\odot$/$V_{24pc}$, as derived by \citep{2014ApJ...781...70F} and \citep{2007ApJ...665L.163D}, respectively. We estimated a lower limit for the cut-off energy on the proton spectrum of $\approx 37$ TeV (at 95\% CL), by comparing the maximum likelihoods of the data obtained for the exponential cutoff power-law and power-law models with the likelihood-ratio test.

The total energy in protons (above 1.22\,GeV) can be compared with the total mechanical power of the stellar winds in the Westerlund 2 cluster:  $\rm{W}_{\rm{tot}}$ =$f L_0T_0$, which results in a modest acceleration efficiency of $f=10^{-4}$ ($10^{-6}$ / $5\times 10^{-3}$), for the well known age ($T_0 =2\times10^6$ yr), a distance of 5\,kpc (2\,kpc / 8\,kpc) and the available energy budget in the form of kinetic energy of stellar winds ($L_0 = 2 \times 10^{38}$ erg s$^{-1}$). 

Even if energetically possible, the uncertainties on the distance of the cluster and the lack of a clear cloud/gamma-ray morphological match make the association unclear. Nevertheless, if assuming that the size observed at the GeV and TeV energies corresponds to the propagation depth of CRs injected by either the cluster or some accelerator within it, the resulting diffusion coefficient obtained from $R_{\gamma} \sim 24\, (d/\rm{5kpc}$) pc, would be of the order of $D \sim 3\times 10^{25} \rm{cm}^2\rm{s^{-1}}$ (or (0.4   - 6)$\times$10$^{25}\rm{cm}^2\rm{s^{-1}}$ for 2 and 8 kpc respectively), which is in tension with the value of the diffusion coefficient at these energies in the Bohm regime  \citep{2019NatAs...3..561A}. That points to a certain CR halo around Westerlund 2, beyond the size traced by LAT and H.E.S.S., that could, in principle, extend up to a few tens of parsecs. However, the uncertainty on the distance to Westerlund\,2 prevents deriving firm conclusions. More precise estimations of the distance to the cluster, foreseen with {\it Gaia} DR3 \citep{2018AJ....156..211Z} (and therefore of the real gamma-ray size), would provide constraining limits on the diffusion of CRs around the source, for a range of acceleration efficiency $f$.

\section{Conclusions}
\label{sec5}

The reanalysis of the large LAT dataset presented in \citep{2021MNRAS.tmp.1423M} results in a clear identification of the extended source \ext\ with the TeV source \hess. The matching spectral and morphological agreement, with no signs of cooling features in the size of the source, points to a common origin of the radiation. The combination of the two results obtained, that is, the extended source beyond the cluster size and in a good agreement with the TeV radiation, and the hard spectrum that continues towards low energies, constitutes evidence of the hadronic nature of the gamma-ray emission detected using \emph{Fermi}-LAT and H.E.S.S. data. Deeper observations with H.E.S.S. or with sensitive TeV instruments in the South such as CTA \citep{2019scta.book.....C} in the future should provide a definitive answer to the PeVatron nature of \hess\ and its connection with the stellar cluster Westerlund\,2.

\end{document}